\documentclass[aps,12pt,pre]{revtex4}

\newcommand{\be}{\beta}
\newcommand{\ps}{pseudospinodal}
\newcommand{\s}{spinodal}
\newcommand{\re}{{\rm Re}}
\newcommand{\im}{{\rm Im}}
\newcommand{\br}{{\re\,\beta}}
\newcommand{\hr}{{\re\,h}}
\newcommand{\bi}{{\im\,\beta}}
\newcommand{\hi}{{\im\,h}}
\newcommand{\ho}{{h_0}}
\newcommand{\tr}{{\re\,T}}

\newcommand{\hs}{h_s}
\newcommand{\mf}{mean-field}

\usepackage{graphicx}
\begin{document}

\title{Zeros of the Partition Function and Pseudospinodals in
Long-Range Ising Models}

\author{Natali Gulbahce}
\affiliation{Clark University, Department of Physics, Worcester,
MA 01610}
\affiliation{Los Alamos National Laboratory, CCS-3, MS-B256, Los
Alamos, NM 87545}

\author{Harvey Gould}
\affiliation{Clark University, Department of Physics, Worcester,
MA 01610}
\author{W. Klein}
\affiliation{Los Alamos National Laboratory, X-7 Material
Science, Los Alamos NM 87545}
\affiliation{Boston University, Department of Physics, Boston, MA
02215}

%\date{\today}

\begin{abstract}
The relation between the zeros of the partition function and
spinodal critical points in Ising models with long-range
interactions is investigated. We find the spinodal is associated with
the zeros of the partition function in four-dimensional complex
temperature/magnetic field space. The zeros approach the real
temperature/magnetic field plane as the range of interaction
increases.
\end{abstract}

\maketitle

\section{\label{sec1}Introduction}
Mean-field treatments of fluids and Ising models yield metastable
and unstable regions, separated by a well-defined line known as
the spinodal~\cite{kac}. As the spinodal is approached, the
system shows phenomena similar to that at mean-field critical
points. In particular, the isothermal susceptibility $\chi$
diverges as a power law as the spinodal value of the magnetic field
$\hs$ is approached from the metastable state~\cite{uk},
\begin{equation}
\label{chi}
\chi \propto (\hs - h)^{-1/2},
\end{equation}
where $h$ is the (dimensionless) magnetic field. A mean-field Ising 
system can be realized in a well defined way by assuming an 
infinite-range interaction between the spins~\cite{kac}. 

If the interaction range is long but finite, the system is no
longer mean-field, but can be described as near-mean-field, and
the spinodal singularity is replaced by a pseudospinodal, which
still has physical effects. As with apparent critical points in
finite size systems, the susceptibility for a finite interaction
range can be fit to a power law over a limited range of the
scaling field
$(\hs - h)$. For example, the susceptibility in the metastable
state of a long but finite-range interaction Ising model appears
to diverge over several decades in $(\hs - h)$ as the
pseudospinodal is approached~\cite{stauffer}. However, the
divergence is suppressed if the
\ps\ is approached too closely, indicating that the spinodal
singularity has been smeared out~\cite{stauffer}. It also is found
that the properties of the \ps\ converge rapidly with increasing
interaction range to those predicted for the spinodal in \mf\
theory~\cite{novotny}.

Because the spinodal is a line of critical points, we expect that the
Ising spinodal has properties similar to those of the
Ising critical point. In particular, we expect that the spinodal is
related to the zeros of the partition function. Lee and
Yang~\cite{leeyang} showed that the singularity in the free energy
for $T$ less than
$T_{c}$, the critical temperature, arises from the presence of a real
positive zero of the partition function in the thermodynamic limit.
In finite systems there is no real positive zero for $T<T_{c}$.

The zeros of the 
partition function for real temperature $T$ lie on the imaginary axis
of the complex magnetic field plane~\cite{latticegas}. For finite
systems there is a gap in the distribution of the zeros around $h=0$.
This gap shrinks to zero for $T<T_{c}$ as $N \rightarrow
\infty$. The Lee-Yang relation between the zeros of the partition
function and critical points is valid for all interaction ranges at the
Ising critical point. This ideas were extended to complex
temperature by Fisher~\cite{fisher}.

In order to generalize these ideas to \s s in Ising models, we
consider an Ising model in a magnetic field $h$. We will consider
the ``infinite-range'' Ising model in which each spin interacts with
all other spins~\cite{milotti}, and the Domb-Dalton~\cite{domb}
version of the Ising model in which each spin interacts with its
neighbors within a given interaction range $R$ with a constant
interaction. These models can be described by mean-field theory in
the limit $N
\to \infty$ and $R\to \infty$ (for $N$ infinite), respectively. Our
main result is that the \ps\ is related to the zeros of the
partition function in four-dimensional complex
temperature/magnetic field space. In addition, the zeros
approach the real temperature/magnetic field plane as the system
becomes more \mf.

The structure of the paper is as follows. In Sec.~\ref{sec2} we
consider the infinite-range Ising model and show both analytically and
numerically that the zeros of the partition function approach the
real $\beta$ and $h$ plane as $N$ increases. We find a similar
result in Sec.~\ref{sec3} by estimating the partition function
in the metastable state by using the Metropolis algorithm and the
single histogram method. In Sec.~\ref{sec4} we consider the
Domb-Dalton version of the Ising model and estimate the
partition function in the metastable state in the same way.

\section{\label{sec2}The infinite-range Ising Model: Analytical
Approach}
We first consider an Ising model in which every spin interacts
with every other spin. We will refer to this model as the
infinite-range Ising model~\cite{milotti}, although the
interaction range becomes infinite only in the limit $N \to
\infty$. The Hamiltonian is
\begin{equation}
\label{eq:ham}
H = -J_N\! \sum_{i \neq j=1}^N
\sigma_i\sigma_j-h\sum_i\sigma_i.
\end{equation}
We need to rescale the interaction so
that the total interaction energy seen by a given spin remains the same as
$N$ is increased~\cite{kac}. We will take
\begin{equation}
\label{J}
J_N = {4J \over N-1}.
\end{equation}
This choice of $J_N$ yields the Ising mean-field
critical temperature $T_c=4$ when $N\to \infty$~\cite{domb}, where we
have chosen units such that $J/k = 1$.

The exact density of states is easily calculated for this model and is
given by
\begin{equation}
\label{eq:gofm}
g(M) = {N! \over n! (N - n)!},
\end{equation}
where $n$ is the number of up spins. We have $M = 2n - N$ and
the total energy $E = J_N(N - M^2)/2$, where $M$ is the
magnetization. (In general, the density of states depends on both
$E$ and $M$, but because $E$ is a unique function of $M$ in the
infinite-range Ising model, we need only write $g(M)$.) Hence
the partition function $Z$ can be expressed analytically for
arbitrary complex inverse temperatures $\beta$ and magnetic fields
$h$:
\begin{equation}
\label{eq:Z}
Z(\beta,\beta h) = \sum_{\rm all\, M} g(M) e^{-\beta E} e^{\beta h
M}.
\end{equation}

To understand the nature of $Z$ in the metastable state, imagine a
simulation of an Ising model in equilibrium with a heat bath at
inverse temperature $\beta$ in the magnetic field
$h=\ho>0$. Because $\ho>0$, the magnetization values will
be positive. Then we let
$h \to -\ho$. If $\ho$ is not too large, the system will remain in
the metastable state for a reasonable amount of time and sample
positive values of
$M$. Hence, to determine the zeros of $Z$ associated with the \ps,
we need to restrict the sum in Eq.~(\ref{eq:Z}) to magnetization
values, $M$, that are representative of the metastable state. The
following examples will illustrate the need for this restriction
and the procedure for determining the zeros of $Z$.

The notion of using a restricted partition function sum to describe
the metastable state has a long history. Penrose and
Lebowitz~\cite{penleb,lebowitz} review such restricted partition
functions and their properties. A more physical approach can be found
in the discussion of the non-interacting droplet model by
Langer~\cite{lang}. In this model fluctuations are restricted to
non-interacting compact droplets of the stable phase occurring in the
metastable phase. The partition function sum is restricted to 
droplets less than the critical size. This approximation is 
reasonable for low temperatures close to the coexistence curve.
Langer showed that this restriction gives the same metastable state
free energy as the analytic continuation of the stable state free
energy in the same model. However, there are additional properties of
the analytic continuation that do not appear in the restricted sum
which are related to the decay of the metastable state rather than the
description of the metastable state itself.

We first consider $N=4$ and retain only the terms in the partition
function sum that correspond to the two positive values of $M$.
{}From Eqs.~(\ref{eq:gofm}) and (\ref{eq:Z}), we have
\begin{equation}
\label{roots}
Z_r(\beta,\beta h) = \sum_{M = 2, 4} g(M) e^{-\beta E} e^{\beta h
M} = e^{8
\beta} e^{-4 \beta h} + 4 e^{-2 \beta h}.
\end{equation}
The subscript $r$ denotes that the sum over $M$ is
restricted. If we let $x = e^{-2 \beta h}$, the equation
$Z_r=0$ is equivalent to,
\begin{equation}
\label{eq:n4}
e^{8 \beta} x + 4 = 0,
\end{equation}
and has the solution
\begin{equation}
\label{eq:n4sol}
\be h = -\ln 2 - i {\pi \over 2} + 4 \beta.
\end{equation}

In general, we have four unknowns (the real and imaginary parts of
$\beta$ and $h$); Eq.~(\ref{eq:n4sol}) yields two
conditions. In the following we will fix
\begin{equation}
\br = 9/16,
\end{equation}
which is equivalent to a temperature of $T={4\over 9}T_c$. 
For this value of
$T$, the value of the spinodal magnetic field is known to be $\hs
\approx 1.2704$~\cite{monette}. Equation~(\ref{eq:n4}) then gives
a line of zeros in complex
$(\beta, \beta h)$ space. However, if we are interested only in the
zero closest to the real 
$\beta$, $\beta h$ plane, we need a fourth condition. This
condition is found by requiring that the
quantity,
\begin{equation}
\label{eq:D2}
D^2 = (\bi)^2 + (\im \beta h)^2,
\end{equation}
be a minimum, which is equivalent to requiring that the leading zero of
$Z_r$, the zero closest to the real $\beta$ and
$\beta h$ plane, be as close to this plane as possible. If we let
$y = \bi$ and use Eq.~(\ref{eq:n4sol}), we can rewrite $D^2$ as
\begin{equation}
D^2 = y^2 + (-{\pi \over 2} + 4 y)^2.
\end{equation}
Because we want $D^2$ to be a minimum, we require
\begin{equation}
{dD^2 \over dy} = 2 y + 2(-{\pi \over 2} + 4y)(4) = 0.
\end{equation}
The solution is
\begin{equation}
\label{12}
y = \bi = {2 \pi \over 17} \approx 0.3696,
\end{equation}
and $D = 0.38097$. Note that $\tr =
\br/(\br^2 +
\bi^2) = 1.2417$. We finally use Eq.~(\ref{eq:n4sol}) to obtain
the value of complex
$h$. The result is summarized in the first row of
Table~\ref{tab:summary}.

The solutions for $N=9,\ 16,\ 100$, and 1000, keeping all the
positive
$M$ contributions to the partition function, also are shown in
Table~\ref{tab:summary}. We see that although $D$ becomes smaller
as $N$ is increased, $|\hr|$ overshoots the mean-field value of
$\hs \approx1.27$ (for $\beta = 9/16$). Hence, retaining all the
positive $M$ terms in the partition function allows the system to
explore more than the metastable state, and we need to further
restrict the sum over values of $M$. Physically we want to exclude
values of $M$ that would drive the system to the stable phase.

\begin{figure}[ht]
\includegraphics{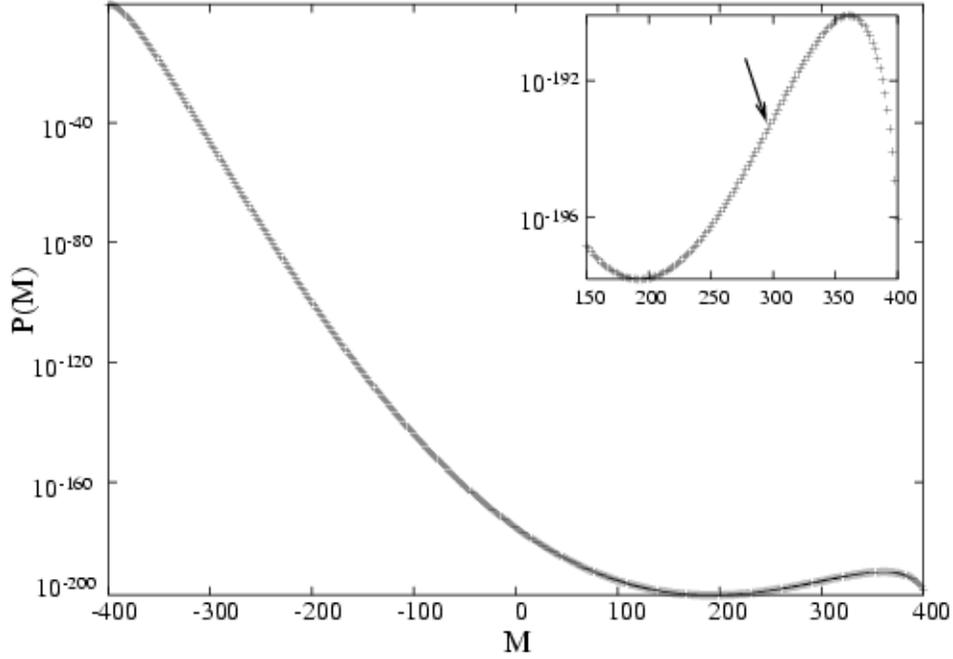}
\caption{\label{fig:pofm400}The probability of the magnetization,
$P(M)$, obtained from Eq.~(\ref{eq:pofm}) for the infinite-range Ising
model with
$N=400$, $h=-1.0$, and $\beta = 9/16$. The values of
$P(M)$ are plotted on a  $\log_{10}$-linear scale because of
the dominance of the negative values of $M$. We include terms in the
partition function sum over $M$ from the inflection point 
of $P(M)$ at $M_I=298$ to
$M=400$ in the partition function. The inset shows the region where
the inflection point is (see arrow).}
\end{figure}

What are the appropriate values of the magnetization that will
keep the system in the metastable state? One way to determine
these values is to look at $P(M)$, the probability that the system
has magnetization $M$ for a particular value of $\beta$
and
$h$:
\begin{equation}
\label{eq:pofm}
P(M) = g(M) e^{-\beta E } e^{\beta h M}.
\end{equation}
Figure~\ref{fig:pofm400} shows $P(M)$ for a system of $N=400$
spins for $h=-1.0$ and
$\beta = 9/16$. The negative magnetization values
have a relatively high probability (because $h<0$)
and correspond to the stable phase. The positive
values of magnetization have a much lower probability, and the
peak at
$M= 360$ corresponds to the most probable value of
$M$ in the metastable state. In between the peak at $M=-400$ and
the peak at $M=360$, $P(M)$ has a minimum at $M_{\min} =
192$ for this value of $h$ and an inflection point at $M_{I} = 298$. 
We will only include values of the
magnetization in the partition function sum that are greater than
$M_I$.

The reason for this choice of the cutoff has to do with the nature of
metastability. We expect that $P(M) \propto \exp(-\beta F(M))$,
where $F(M)$ is the metastable state free energy. We
want to exclude from the partition function values of $M$ that 
correspond to states which are not characteristic of 
equilibrium. In the infinite range model 
the configurations with $M_{\min} < M < M_I$ are unstable in 
that the initial evolution of a fluctuation does not monotonically 
decay to the metastable well. This behavior
translates into an initial growth of fluctuations rather than the
monotonic decay expected in equilibrium.

This behavior is a consequence
of the fact that the free energy, which is proportional to $\log(P(M))$,
is not convex for these values of $M$. Obviously, this behavior holds
for a range of values of $M < M_{\min}$ as well. However, we can
exclude all configurations with 
$M < M_{\min}$ because they are in the stable free energy
well and do not occur in the metastable state.
Our particular choice of the cutoff is well defined, but is arbitrary to
some extent as long as
$M$ is greater than $M_I$. 

To determine $M_{I}$, we calculate the second derivative 
of $P(M)$ as given in Eq.~(\ref{eq:pofm}).
We find the value of $M$ that satisfies
\begin{equation}
\label{inflect}
{\partial^{2} P(M)\over \partial M^{2}} = {-N \over (N-M)(N+M)} + \beta
J = 0.
\end{equation}
Clearly there will be two inflection
points (see Fig.~\ref{fig:pofm400}). We choose the one closest to the
metastable state maximum of $P(M)$. 
We find that the value of $M_{I}$ is independent
of $h$, which is consistent with 
the idea that the free energy for this system can be written in the
Landau-Ginzburg form where the magnetic field appears only in a term
linear in $M$.

We now write the restricted
partition function $Z_r(\beta,\beta h)$ as 
\begin{equation}
\label{eq:Zr1}
Z_r(\beta,\beta h) = \sum_{M=M_I}^N C_M \,x^{M/2},
\end{equation}
The coefficients, $C_M$, extend
over a wide range of values and are as large as $10^{200}$ for the
values of $N$ that we considered. For this reason we computed
$C_M$ to arbitrary precision so as not to lose accuracy. The zeros
of
$Z_r$, which is a polynomial in
$x$, were found using MPSolve~\cite{bini,mpsolve}.

\begin{figure}[ht]
\includegraphics{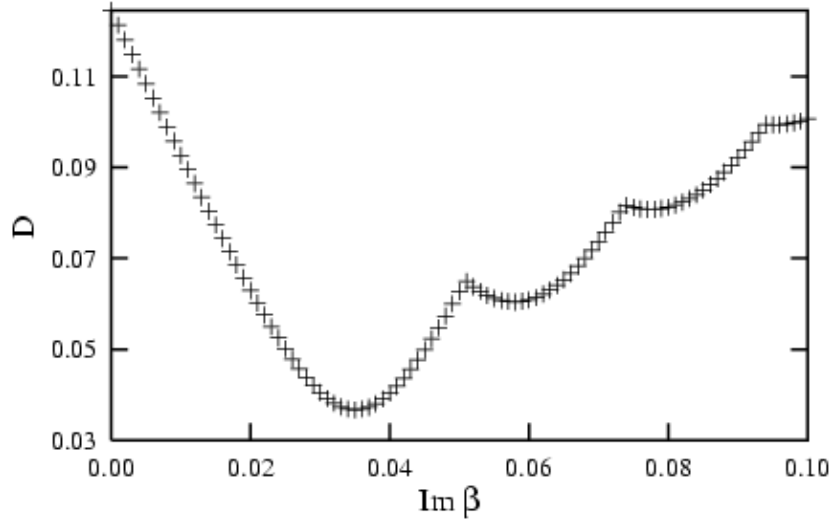}
\caption{\label{fig:dvsalpha} The value of $D$, a measure
of the distance of the leading zero to the real $\beta$ and $\beta
h$ plane, versus
$\bi$ for the infinite-range Ising model for $N=400$. 
The minimum distance to the real axes occurs at $\bi \approx
0.035$ for this value of $N$.}
\end{figure}

For a given value
of $\bi$, we solve for the zeros of
$Z_r$ in Eq.~(\ref{eq:Zr1}) and find the value of $x$ that
corresponds to the leading zero, the zero that minimizes $D$ in
Eq.~(\ref{eq:D2}). We repeat this step for a range of values of
$\bi$ and determine numerically the value of
$\bi$ that yields the minimum value of $D$. 
The typical dependence of $D$ on $\bi$ is shown in
Fig.~\ref{fig:dvsalpha}. From Fig.~\ref{fig:dvsalpha} we see that
for $N=400$, $D$ is a minimum for $\bi
\approx 0.035$. Once we know this value of $\bi$, we solve for $h$
from the relation $h = \log x/(-2\beta)$. (The value of
$x$ was determined from the solution of $Z_r=0$.)

\begin{figure}[ht]
\includegraphics{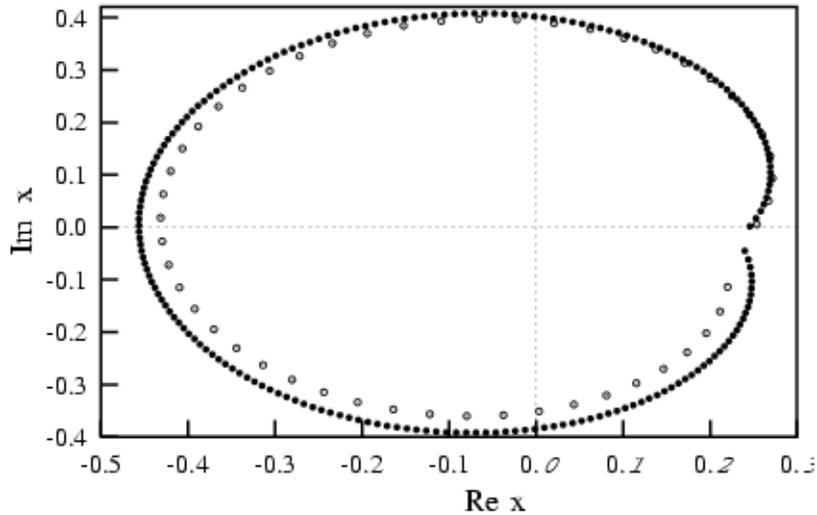}
\caption{\label{fig:roots} The imaginary versus the real part
of the zeros of the partition function plotted in terms of the
variable $x=e^{-2\beta h}$ for $N=400$ (empty circles) and 1600
(filled circles) for the infinite-range Ising model. The zeros
were obtained by the analytical method described in
Sec.~\ref{sec2} using $\bi$ listed in Table~\ref{tab:analres}.}
\end{figure}

\begin{figure}[ht]
\includegraphics{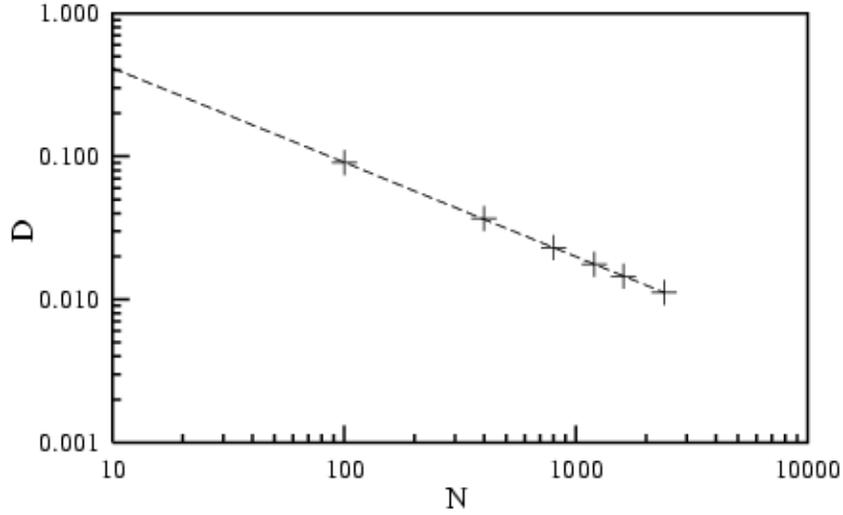}
\caption{\label{fig:dvsNanalitik} Log-log (base 10) plot of $D$, the distance 
of the leading zero of the partition function to
the real $\beta$-$\beta h$ axes (see Eq.~(\ref{eq:D2})), versus the
system size
$N$, using the analytical approach described in Sec.~\ref{sec2}. The slope is
$-0.659
\pm 0.003$. The data is from Table~\ref{tab:analres}.}
\end{figure}

We repeat the above steps for a range of values of $N$
and obtain $D$, $\bi$, $\hr$, and $\hi$. 
Our
results are summarized in Table~\ref{tab:analres}. Note that
$\hi$, $\bi$, and $D$ decrease as $N$ increases and $|\hr|$
approaches $\hs=1.27$. A plot of the zeros of $Z$ for the
infinite-range Ising model in the $\im\,x$, $\re\,x$ plane is shown
in Fig.~\ref{fig:roots}. The values of $D$ listed in
Table~\ref{tab:analres} are plotted as a function of $N$ in
Fig.~\ref{fig:dvsNanalitik}. Because this log-log plot indicates a
power-law dependence, we write
\begin{equation}
\label{eq:a}
D \propto N^{-a}.
\end{equation}
A least squares fit gives
$a=0.659\pm 0.003$. The estimate of the error is only
statistical.

Our numerical result for the exponent $a$ can be understood by a
simple scaling argument. In order for a mean-field approach,
including the idea of a spinodal, to be a reasonable
approximation, the system must satisfy the Ginzburg criterion,
that is, the Ginzburg parameter $G$ must be much greater than
unity. For the infinite-range Ising model, the Ginzburg criterion
can be written as~\cite{landau}:
\begin{equation}
\label{ginz1}
{\xi^{d}\chi \over \xi^{2d}\phi^{2}} << 1,
\end{equation}
where $\xi$ is the correlation length, $\phi$ is the order
parameter, and $d$ is the spatial dimension. We have $\chi
\sim {\Delta h}^{-1/2}$ and
$\phi
\sim {\Delta h}^{1/2}$~\cite{uk}, where $\Delta h = h_{s} - h$, to
obtain
\begin{equation} 
\label{ginz2}
G = \xi^{d} {\Delta h}^{3/2} >> 1.
\end{equation}
For the infinite-range model, $N \sim {\xi}^{d}$. Hence,
\begin{equation}
\label{eq:ginzburg}
G = N \Delta h^{3/2},
\end{equation}
up to a constant of order unity.
The Ginzburg parameter $G$ is a measure of how mean-field the
system is for finite
$N$; the larger the value of $G$, the more mean-field the system
is. If we keep $G$ constant as we approach the spinodal, we see
that
\begin{equation}
\label{eq:deltaH}
\Delta h \propto N ^{-2/3}.
\end{equation}
We present an argument for why $G$ should be held constant in
Sec.~\ref{discuss}.

{}From Eq.~(\ref{eq:deltaH}) we see that $\Delta h$ and the
distance $D$ approach zero with an exponent $a=2/3$, in good agreement with
our numerical result. Because $\Delta h$ in Eq.~(\ref{eq:deltaH}) could be
associated with $\hi$, $(\hs - |\hr|)$, or $D$, we note that $\hi$ in
Table~\ref{tab:analres} goes to zero with the same exponent as in Eq.~(\ref{eq:a}).
We also find that $(\hs - |\hr|)
\propto N^{-0.61}$. 

\section{\label{sec3}The infinite-range Ising Model: 
Monte Carlo Approach}

In general, the partition function is not known analytically.
However, we can use a Monte Carlo (MC) method to determine the 
density of states from which we can determine an estimate of the 
partition function. Such an approach has been used to find the
density of states for the nearest-neighbor Ising
model~\cite{bhanot, alves}. In this way the leading partition
function zeros at the critical point have been computed and the
critical exponent $\nu$ and corrections to scaling have been found
with high precision~\cite{alves}. Our goal is not to obtain
precise estimates of the critical exponents near the \ps, but to
show that the same simulations that show an apparent divergence in
the susceptibility also yield an estimate for the leading zero of
the partition function which behaves as expected as the
system becomes more mean-field.

\begin{figure}[ht]
\includegraphics{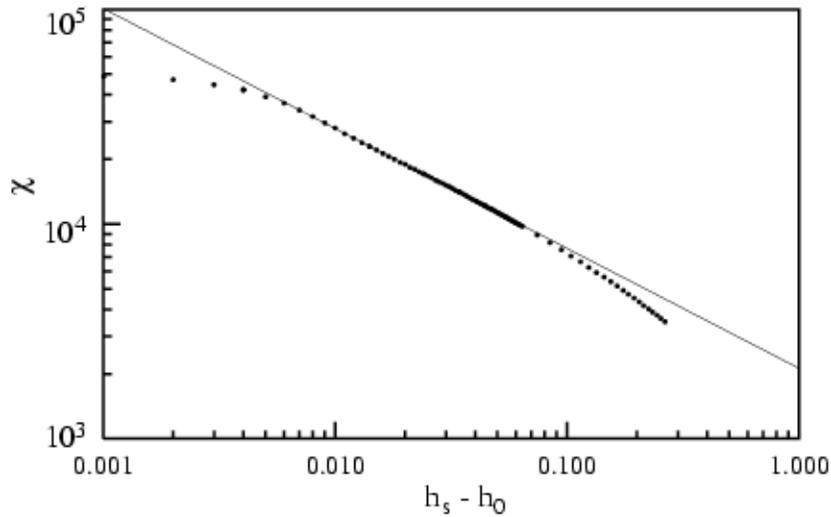}
\caption{\label{fig:xvsh} Log-log (base 10) plot of the
isothermal susceptibility
$\chi$ as a function of
$(\hs - \ho)$ for the infinite-range Ising model with 
$N=10000$. The
system was equilibrated using the Metropolis algorithm at a
temperature 
$T=16/9$ and applied field $h=\ho$. Then the field was flipped and
the values of the magnetization were sampled in the metastable
state. Note that $\chi$ exhibits mean-field behavior over about 2
decades and the apparent divergence of $\chi$ at the spinodal
field $\hs=1.27$ is rounded off when $(\hs - \ho)$ becomes too
small. This behavior is an example of the influence of a \ps. The
straight line with a slope of 1/2 is drawn as a guide to the eye.}
\end{figure}

To this end we use the Metropolis algorithm to equilibrate the
system at temperature $T=16/9$ and 
applied magnetic field $h=\ho$ for about 100 MC
steps per spin. (The system equilibriates as quickly as 10
MC
steps per spin depending on the strength of the field.) Then we
flip the magnetic field and compute the histogram $H(E,M)$ from
which we determine the densify of states $g(E,M)$ and the
partition function for complex
$\beta$ and $h$. We save the values
of $M$ after $h \to -\ho$ and run until $M$ changes sign or for
5000 MC steps per spin, whichever comes first. We then throw away
the first 20\% of the data to ensure that the system is in
metastable equilibrium and the last 20\% of the data to ensure that
we do not retain values of $M$ that are too close to the stable
state. The remaining 60\% of the run is used to obtain
$H(E,M)$. We also omit any run whose lifetime in
the metastable state is less than 100 MC steps per spin. Our
results for
$H(E,M)$ are not sensitive to the choice of the minimum lifetime
nor the percentage of each run that we use to estimate
$H(E,M)$. (Because
$E$ is a function of $M$ for the infinite-range Ising model, we
need only to compute $H(M)$. However, we need to compute $H(E,M)$ in
Sec.~IV.) We averaged
$H(E,M)$ over approximately 5000 runs for each value of
$\ho$ for a total of approximately $1.5 \times 10^7$\,MC steps per
spin for a given value of $h$ and $N$. Our results for the
susceptibility $\chi$ are given in Fig.~\ref{fig:xvsh}. As
mentioned in Sec.~\ref{sec1}, $\chi$ shows an apparent divergence
with an mean-field exponent of 1/2 until the \ps\ is
approached too closely.

Given the histogram $H(E,M)$ at 
$\beta_0=9/16$ and $h=-\ho$, we use the usual single histogram
method~\cite{ferrenberg} and express the partition function for
arbitrary (complex) $\beta$ and
$h$ as:
\begin{equation}
\label{eq:Zw}
Z_m(\beta,\beta h) = \sum_{E,M} H(E,M) e^{(\beta_0 -\beta) E}
e^{-(\beta_0\ho - \beta h) M}.
\end{equation}
Note that we do not have to determine the lower cutoff for $M$
because the Monte Carlo simulation only samples values of $M$
while the system is in a metastable state (noted by the subscript
$m$ in Eq.~(\ref{eq:Zw})). As $\ho$ is
increased for fixed $N$, the distance $D$ initially decreases,
but then begins to increase as 
$\hs$ is approached too closely. That is, $D$ shows a minimum as
a function of $\ho$. For each of value of $N$ we choose the value
of $\ho$ for which $D$ is a minimum. 
A comparison of the
histogram that was determined analytically in Sec.~\ref{sec2} and
estimated by the Metropolis Monte Carlo algorithm shows similar
qualitative behavior, except that the latter is approximately a 
Gaussian and extends to lower values of $M$ (see
Table~\ref{tab:mcres}), but with a smaller amplitude. [xx note change
xx]

\begin{figure}[ht]
\includegraphics{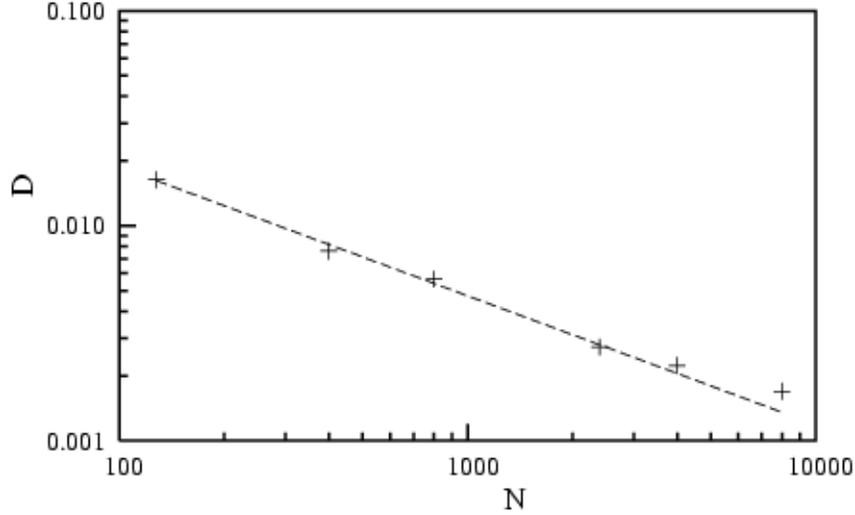}
\caption{\label{fig:dvsnmc} Log-log (base 10) plot of $D$, the distance of
the leading zero of the partition function to the real axes, 
versus the system size
$N$ for the infinite-range Ising model obtained from Monte Carlo
simulations. A least
squares fit gives a slope of $-0.60 \pm 0.03$. The data is from
Table~\ref{tab:mcres}.}
\end{figure}

\begin{figure}[ht]
\includegraphics{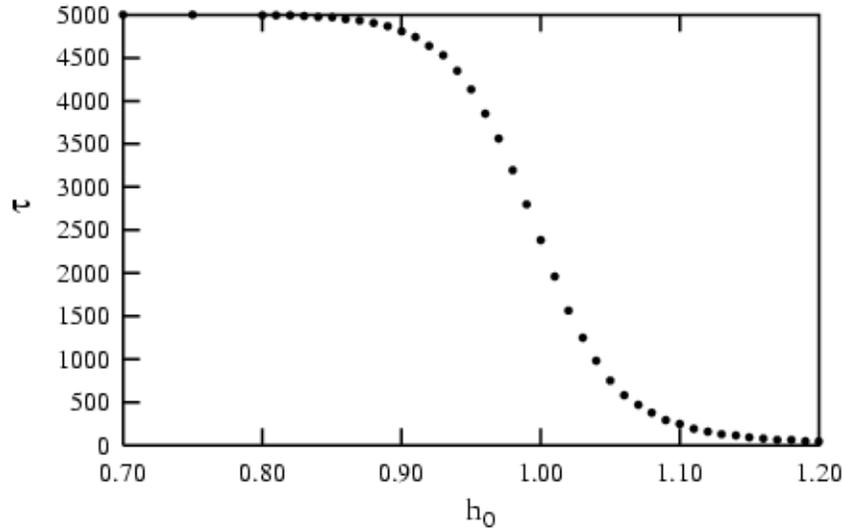}
\caption{\label{fig:lifetime} The lifetime $\tau$ of the metastable
state in the infinite-range Ising model as a function of the
applied field
$\ho$ for
$N=128$. The behavior of $\tau$ for the long-range Ising model
considered in Sec.~\ref{sec4} is similar, and for the latter we
choose
$\ho$ to be the field at which $\tau(\ho)$ just begins to decrease 
sharply.} 
\end{figure}

Table~\ref{tab:mcres} shows our results for $h$, $\bi$, and
$D$ for a range of values of 
$N$ at the values of $\ho$ that minimize the distance $D$. A
log-log plot of
$D$ versus
$N$ for the data in Table~\ref{tab:mcres} is shown in
Fig.~\ref{fig:dvsnmc}; a least squares fit gives $a=0.60 \pm
0.03$, which is consistent with the result obtained using the exact
density of states (with a cutoff). Note that $\tau$, the
average lifetime of the metastable state, for each value of $N$ is
approximately a constant at the value of $\ho$ that was chosen to
minimize
$D$ (see Fig.~\ref{fig:lifetime}). We will use this fact to
choose the value of $\ho$ for the long-range Ising model in
Sec.~\ref{sec4}.

\section{\label{sec4}Long-range two-dimensional Ising
model}

As discussed in Sec.~\ref{sec1}, the susceptibility of
long-range Ising models in the metastable state shows
an apparent divergence as the applied magnetic field is increased.
In the following we show that this effect of a \ps\ in long-range
Ising models is reflected in the behavior of the zeros of the
partition function as a function of the complex temperature and
magnetic field. We will show that as the interaction range $R$
increases, the leading zero moves closer to the real plane.

Following Refs.~\onlinecite{kac} and \onlinecite{domb}, we consider
an Ising model such that each spin interacts with its neighbors
within a given interaction range
$R$ with a constant interaction $J=4/q$, where $q$ is the number
of interaction neighbors. (The factor 4 is included so that $J=1$
for the usual Ising model on the square lattice.)
If the thermodynamic limit is taken
first~\cite{kac}, the system is mean-field in the
$R\rightarrow \infty$ limit, and the system is described by 
Curie-Weiss theory~\cite{lebowitz}.
In this limit the metastable state ends at a spinodal point. The
spinodal is a critical point and the susceptibility $\chi$
diverges as in Eq.~(\ref{chi}).
 
We consider the Ising system on a square lattice with linear
dimension $L=240$ and
$N= 57600$. The interaction range $R$ is defined such that a given
spin interacts with any spin that is within a circle of radius
$R$. The number of neighbors of a given spin is shown in the second
column of Table~\ref{tab:resultsrange}. This system is large enough
($L$ is 10 times larger than the maximum value of $R$) for finite
size effects to be minimal, but the finite size of the system
implies that the zeros of the partition function must be
complex for any range $R$.

As in Sec.~\ref{sec3}, we equilibrate the system at
inverse temperature $\beta = 9/16$ and applied magnetic field
$\ho$ for 100 MC steps per spin. Then we flip the field and compute
the histogram
$H(E,M)$ from which we determine the density of states $g(E,M)$
and the partition function for complex
$\beta$ and $h$. The most time consuming part of our procedure is
determining the appropriate value of $\ho$. Although we could
determine $\ho$ by minimizing $D$ as in Sec.~\ref{sec3},
we instead determined the mean lifetime $\tau$ of the metastable
state as a function of $\ho$. For small $\ho$ (away from the
spinodal),
$\tau$ is greater than the duration of our runs which
are $2 \times 10^4$ Monte Carlo steps per spin. As
$\ho$ approaches
$\hs$, the lifetime begins to decrease. As mentioned in
Sec.~\ref{sec3}, we found that the values of $\ho$ that minimize
$D$ are near $\hs$ and 
also yield approximately constant lifetimes for different values of $N$.
(This behavior was found for both the infinite- and long-range
Ising models.) We choose the largest value of
$\ho$ for which the mean lifetime of the metastable state just
begins to decrease below $2
\times 10^4$ Monte Carlo steps per spin. This criterion for
choosing
$\ho$ is not as sensitive as minimizing
$D$, but is much quicker although some additional error is
introduced when $\ho$ is chosen this way. For a given $\ho$,
$10^3$ runs of
$2 \times 10^4$ MC steps per spin were done to determine the
histogram $H(E,M)$. 

\begin{figure}[ht]
\includegraphics{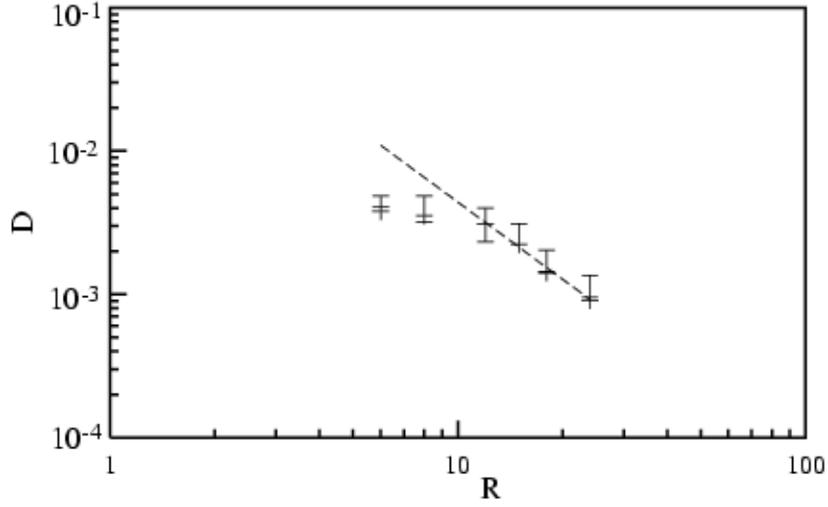}
\caption{\label{fig:dvsr} Log-log (base 10) plot of the distance $D$ to the
real $\beta$-$\beta h$ axes versus the interaction range $R$ for
the two-dimensional Ising model. Note that the leading zero of the
partition function moves closer to the real axes as the system 
becomes more mean-field. A least squares fit gives a slope of
$-1.8 \pm 0.3$. The data is from Table~\ref{tab:resultsrange}.}
\end{figure}

\begin{figure}[ht]
\includegraphics{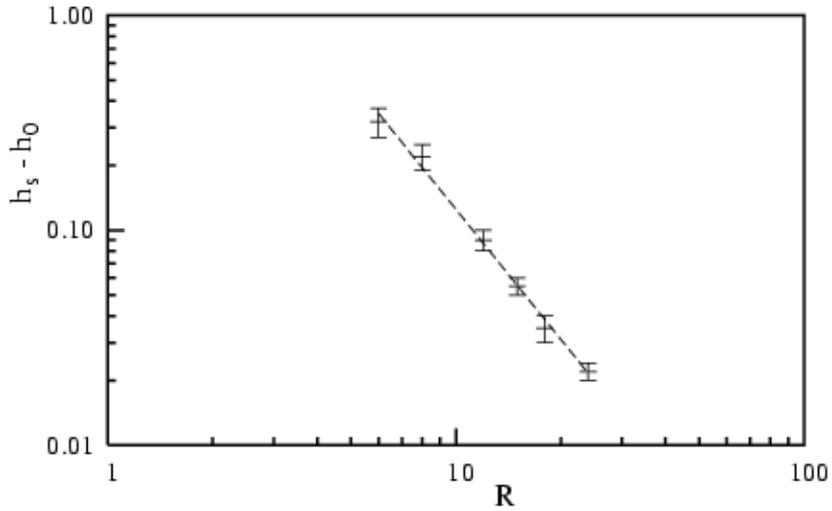}
\caption{\label{fig:h0vsr} Log-log (base 10) plot of the difference
$\hs-\ho$ as a function of the interaction range $R$. A least
squares fit gives a slope of $-2.01 \pm 0.07$. The data is from
Table~\ref{tab:resultsrange}.}
\end{figure}

The resulting values of the distance $D$ as a function of the
interaction range
$R$ are listed in Table~\ref{tab:resultsrange} and are plotted in
Fig.~\ref{fig:dvsr}. We find 
\begin{equation}
\label{eq:Dlr}
D \propto R^{-1.8 \pm 0.3} .
\end{equation}
The largest contribution to the above error estimate arises
from the uncertainty in the values of the applied fields $\ho$.
Similarly, we find that the difference $(\hs-\ho)$ varies with $R$
as (see Fig.~\ref{fig:h0vsr})
\begin{equation}
\label{eq:holr}
\hs-\ho \propto R^{-1.97 \pm 0.06} .
\end{equation}
The scaling behavior of $(\hs -|\hr|)$ is similar.
The systematic error due to the uncertainty in $\ho$ is
the largest contribution to the error estimates.

The scaling behavior of $D$ and
$\hs-\ho$ can be understood by a scaling argument similar to that
given in Sec.~\ref{sec2}. For a finite range system, the Ginzburg
parameter is~\cite{monette}
\begin{equation}
\label{eq:ginzburgr}
G = R^d \Delta h^{3/2-d/4} .
\end{equation}
For two dimensions, Eq.~(\ref{eq:ginzburgr}) becomes
\begin{equation}
\label{eq:5}
G = R^2 \Delta h .
\end{equation}
Hence, if we keep the Ginzburg parameter fixed, we conclude that
\begin{equation}
\Delta h \propto R^{-2},
\end{equation}
which is consistent with Eqs.~(\ref{eq:Dlr}) and (\ref{eq:holr}).

\section{Discussion}\label{discuss}
We have shown that the \ps\ in Ising models has a well defined
thermodynamic interpretation and can be associated with the
leading zero of the partition function in complex
temperature/magnetic field space in analogy with the behavior of
the Ising critical point in finite systems. Our results for the
nature of the approach of the leading zero of the partition
function to the real temperature and magnetic field plane are
consistent with simple scaling arguments.

An essential ingredient in the scaling arguments was the condition 
that the Ginzburg parameter $G$ was held constant as the system
approached the \ps. As was seen in Sec.~III, choosing the value
of $\ho$ that minimizes the distance of the leading zero to the
real temperature/magnetic field plane also leads to a
metastable state lifetime that is found numerically to be
constant. From nucleation theory near the spinodal, we know that
the lifetime of the metastable state,
$\tau$, is given by~\cite{uk},
\begin{equation}
\label{life}
\tau\propto {{\Delta h}^{1/2} e^{G}\over R^{d}{\Delta h}^{-d/4}} ,
\end{equation}
where $G$ is defined in Eq.~(\ref{eq:ginzburgr}). For large $G$ and
$R$ as well as small $\Delta h$ (close to the \ps), it is easily
shown that constant $\tau$ implies constant $G$. To see this relation
we simply replace $R$ by $R + \delta R$ and $\Delta h$ by $\Delta h +
\delta h$ in Eq.~(\ref{life}) and demand that $\tau$ remain
constant. For 
$\delta R$ and $\delta h$ small, constant $\tau$ implies constant
$G$. Because $G$ is constant, the scaling arguments of Secs.~III
and IV follow. 

The relation between the zeroes of the partition function and the
spinodal provides a mathematical foundation for the notion of a
\ps\ and clarifies the extent to which spinodals act
like critical points. It also provides a possible route by which
\ps s in supercooled liquids can be
characterized~\cite{klein}. 

\begin{acknowledgments}
We would like to thank Greg Johnson and Frank Alexander for very useful
discussions. This work was carried out in part at Los Alamos National Laboratory 
(LA-UR 03-5959) under the auspices of the Department of Energy and supported by
LDRD-2001501 and was supported in part by the National Science Foundation
under grants PHY99-07949 and PHY98-01878.

\end{acknowledgments}

\clearpage
\newpage

\section*{Tables}

\begin{table}[ht]
\caption{\label{tab:summary}Results for the
infinite-range Ising model if all positive values of
$M$ are retained in $Z$. For larger $N$, $|\hr|$
overshoots $\hs=1.27$ and goes to zero as $N$ is increased still
further.}
\begin{center}
\begin{ruledtabular}
\begin{tabular}{rcccc}
$N$ & $\bi$ & $|\hr|$ & $\hi$ & $D$ \\
\hline
4 & 0.3696 & $1.8577$ & 1.3849 & 0.3810 \\
9 & 0.1704 & $1.1987$ & 0.4782 & 0.1823 \\
16 & 0.1070 & $1.1359$ & 0.2926 & 0.1153 \\
100 & 0.0145 & $0.8854$ & 0.0366 & 0.0164 \\
1000 & 0.0014 & $0.8335$ & 0.0035 & 0.0016 \\
\end{tabular}
\end{ruledtabular}
\end{center}
\end{table}

\begin{table}[ht]
\caption{\label{tab:analres}Values of $M_I/N$, $|\hr|$,
$\hi$, $\bi$ and the distance $D$ for increasing values of $N$ for
the infinite-range Ising model. As explained in the text, the
inflection point of P(M) determines $M_I$, the cutoff for $M$. 
Note that $|\hr|$ approaches $\hs=1.27$ and $M_I/N$ approaches $m_s =
0.745356$. For $N=4000000$, $M_I/N$ = 0.745356.}
\begin{ruledtabular}
\begin{tabular}{rcccccc}
$N$ & $M_I/N$ & $\bi$ & $|\hr|$ & $\hi$ & $D$ \\
\hline
\hline
100 &0.7400 &0.086 &1.1601 & 0.2257 & 0.0902\\
400 &0.7450 &0.035 &1.2125 & 0.0949 & 0.0367\\
800 &0.7450 &0.022 &1.2322 & 0.0603 & 0.0230\\
1200 &0.7450 &0.017 &1.2407 & 0.0453 & 0.0176\\ 
1600 &0.7450 &0.014 &1.2456 & 0.0375 & 0.0145\\
2400 &0.7458 &0.011 &1.2508 & 0.0280 & 0.0112\\
\end{tabular}
\end{ruledtabular}
\end{table}

\begin{table}[ht]
\caption{\label{tab:mcres}Results from Monte
Carlo simulations of the infinite-range Ising
model. The simulations were done in the applied
field $-\ho$ and at the inverse temperature 9/16. The values
of $\ho$ for each value of $N$ were chosen so that
the distance $D$ to the real $\beta$-$\beta h$ plane is a minimum. As
noted in the text, this criterion for
$\ho$ also yields metastable state lifetimes that are roughly
constant for the different values of $N$. The values of $M_{\rm cut}$
represent the smallest values of $M$ that were sampled in the metastable
state.}
\begin{ruledtabular}
\begin{tabular}{rlcrcccc}
N & $\ho$ & $M_{\rm cut}/N$ & $\tau$ & $\bi$ & $|\hr|$ & $\hi$ & $D$
\\\hline
\hline
128 & 0.9  &0.23 & 4808  & 0.0144 & 1.181 & 0.0165& 0.0163 \\
400 & 1.0  &0.44 & $>5000$ & 0.0070 & 1.196 & 0.0095& 0.0076 \\
800 & 1.1  &0.56 & $>5000$ & 0.0050 & 1.229 & 0.0063& 0.0056 \\
2400 & 1.205 &0.59 & 4995  & 0.0026 & 1.256 & 0.0043& 0.0027 \\
4000 & 1.226 &0.63 & 4992  & 0.0021 & 1.264 & 0.0033& 0.0022 \\
8000 & 1.246 &0.65 & 4908  & 0.0016 & 1.269 & 0.0024& 0.0017 \\
\end{tabular}
\end{ruledtabular}
\end{table}

\begin{table}[ht]
\caption{\label{tab:resultsrange}Summary of results for the
Ising model with interaction range $R$ on the square lattice with
linear dimension
$L=240$. The number of neighbors $q$ within an interaction
range $R$ is given in the second column. The value of $\ho$ is
determined for each $R$ by choosing the lifetime of the metastable
state to be approximately $2 \times 10^4$ MC steps per spin. The
duration of each run was $2 \times 10^4$ MC steps per spin and
each run was repeated $10^3$ times. Because we did not use the
first and last 20\% of each run, the total number of MC steps per
spin for each value of $R$ was $1.2 \times 10^7$.}
\begin{ruledtabular}
\begin{tabular}{rrllccc}
$R$ & $q$ & $\ho$ & $\bi$ & $|\hr|$ & $\hi$ & $D$ \\
\hline
6 & 112 & 0.95 & 0.0036 & 0.992 & 0.0084 & 0.0038\\
8 & 196 & 1.05 & 0.0035 & 1.011 & 0.0071 & 0.0035\\
12 & 440 & 1.18 & 0.003 & 1.199 & 0.0078 & 0.0031\\
15 & 708 & 1.215 & 0.0025 & 1.230 & 0.0055 & 0.0026\\
18 & 1008 & 1.235 & 0.0013 & 1.252 &0.0039 & 0.0014\\
24 & 1792 & 1.248 & 0.00095& 1.259 &0.0022 & 0.0010\\
\end{tabular}
\end{ruledtabular}
\end{table}

\clearpage
\newpage

\section*{Figures}

\end{document}